# High Open Circuit Voltage Solar Cells based on bright mixed-halide CsPbBrI$_2$ Perovskite Nanocrystals Synthesized in Ambient Air Conditions

*Sotirios Christodoulou†‡, Francesco Di Stasio†‡, Santanu Pradhan†, Alexandros Stavrinadis†, Gerasimos Konstantatos\*†§*

† ICFO-Institut de Ciencies Fotoniques, The Barcelona Institute of Science and Technology, 08860 Castelldefels (Barcelona), Spain

§ ICREA—Institució Catalana de Recerca i Estudis Avançats, Passeig Lluís Companys 23, 08010 Barcelona, Spain

Corresponding Author

*gerasimos.konstantatos@icfo.eu

**ABSTRACT**

Lead halide perovskite nanocrystals (NCs) are currently emerging as one of the most interesting solution processed semiconductors since they possess high photoluminescence quantum yield (PLQY), and colour tunability through anion exchange reactions or quantum confinement. Here, we show efficient solar cells based on mixed halide ($CsPbBrI_2$) NCs obtained via anion exchange reactions in ambient conditions. We performed anion exchange reactions in concentrated NC solutions with $I^-$, thus inducing a PL red-shift up to 676 nm, and obtaining a high PLQY in film (65%). Solar cell devices operating in the wavelength range 350-660 nm were fabricated in air with two different deposition methods. The solar cells display a photo-conversion efficiency of 5.3% and open circuit voltage (Voc) up to 1.31V, among the highest reported for perovskite based solar cells with band gap below 2eV, clearly demonstrating the potential of this material.

**TOC GRAPHICS**

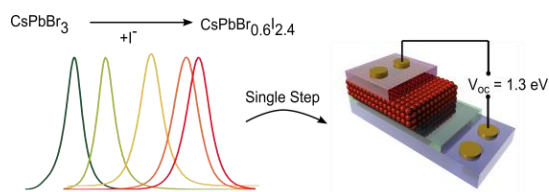

# INTRODUCTION

Inorganic[1] and hybrid lead halide perovskites nanocrystals[2] (NCs) are currently under the spotlight for a variety of optoelectronic applications such as light-emitting diodes,[3–5] solar cells,[6,7] white phosphors[8,9] and solar concentrators.[10] A variety of different methods for the synthesis of $APbX_3$ (where A = Cs, $CH_3NH_3$ or $CH(NH_2)_2$ and X = Br, Cl, I) perovskite[11,12] NCs have been reported in the last 2 years, enabling fabrication of different shapes such as nanocubes,[1] nanowires[13,14] and nanoplatelets.[15,16] Perovskites based on $I^-$ are by far the most interesting for optoelectronic applications, as their absorption spectra extend over the whole visible spectrum. In fact, solar cells based on hybrid bulk iodide perovskite have shown efficiencies up to 22.1%,[17] (certified up to 22.7%) and they can serve as a high band-gap material for tandem solar cells reaching high photo-conversion efficiencies ( >25%).[18–20] The realization of even higher performance tandem solar cells would require a larger bandgap solar cell technology (of approximately 2 eV) that can deliver high $V_{oc}$, in order to complement efficiently with other established thin film or silicon PV technologies. However, bulk inorganic perovskite materials suffer from phase instability in ambient conditions.[7] This unstable cubic crystal structure of $CsPbI_3$ stimulated investigation on its stabilization, which has been achieved by synthesizing the material in nanocrystalline form. Swarnkar et al. succeeded in stabilizing the cubic phase of $CsPbI_3$ NCs via a multiple purification procedure employing an anti-solvent[7] and they have been able to fabricate an efficient solar cell using a layer-by-layer deposition approach. On the other hand, only the direct synthesis of iodide based NCs allows the formation of high quality material, which requires inert atmosphere and high temperatures.[1] Room temperature (RT) and ambient processing are very important factors towards large-scale synthesis and high throughput manufacturing, both needed for large-area and low-cost applications. So far, most of the RT synthetic approaches for CsPbI$_3$ NCs have shown relatively low photoluminescence quantum yield (PLQY): from 20% to 40% in solution.[21,22,23] Only few examples have shown relative high PLQY in solution synthesized at low temperatures.[24,25] Nevertheless, the enhanced PLQY in a diluted solution where the NC are well passivated, it is not always maintained in a dry film, where energy transfer, poor passivation and moisture exposure might take place. Here, we report the preparation of bright and stable mixed halide CsPbBr$_x$I$_{3-x}$ NCs via anion exchange in ambient conditions. The high material quality is manifested by a PLQY of 65% in a dry film. Moreover, we show that the fast and room temperature synthesis can take place in concentrated solutions, thus paving the way to large scale synthesis. The obtained NCs enable the fabrication of solar cells with high open circuit voltage (V$_{oc}$), leading to low-cost, solution processed large-bandgap photovoltaics.

**EXPERIMENTAL SECTION**

**Chemicals**: Lead(II) bromide (PbBr2, 99.999% trace metals basis), Lead(II) iodide (PbI2, 99.999% trace metals basis), Octane (synthesis grade), Oleic Acid (OlAc, > 99%), Caesium acetate (CsAc, 99.99% trace metals basis), Octylamine (OcAm, 99%), Octanoic acid (OcAc, 98%), Propionic acid (> 99.5%), Toluene (TOL, 99.8%) and Methyl acetate (anhydrous, >98%) were purchased from Sigma-Aldrich. 1-propanol (PrOH, Pharmpur®) and n-Hexane (Hex, 99%) were purchased from Sharlab. All chemicals were used without any further purification.

**Synthesis and purification of CsPbBr$_3$ nanocrystals**: CsPbBr$_3$ NCs were prepared following our previously published method according to the ref 27 of the manuscript.

**Anion and ligand exchange reaction**: In 2 ml of concentrated (120 mg/ml) $CsPbBr_3$ NCs dispersed in toluene, 2 ml solution (230 mg/ml) of $PbI_2$ diluted in a mixture of octylamine, octanoic acid and hexane (1:1:1) was added dropwise while gently stirred. The solution turned from green to orange and finally red. After the $PbI_2$ addition, the solution was further stirred for several minutes. Then, the perovskite NCs were collected with centrifugation (1000 RPM for 2 min ) without using any anti-solvent. The NCs were dispersed in 2 ml of toluene and 2 ml of Methyl acetate was introduced to precipitate the NCs and remove the excess ligands. Finally the NCs were disperse in 1 ml of Octane. For ligand exchange reactions, we substitute the octylamine/octanoic acid mixture with the desirable primary alkyl carboxylic acid/amine.

**Device fabrication**: The FTO substrates were sonicated with acetone and propanol for 20 minutes. Then the FTO surface was further clean with oxygen plasma. On FTO glass a sol-gel solution of TiO2 was spin coated at 3000 RPM for 30 sec resulting in a 50 nm layer. The substrates were annealed for 1 h at 500 ºC. The sol-gel $TiO_2$ was prepared according to the ref. 7 of the manuscript. For the single step device a concentrated solution (120 mg/ml) of $CsPbBrI_2$ NCs dispersed in octane was spin-cast at 2000 RPM for 30sec. Then 5 drops of saturated Pb(OMeAc) solution cover the film for 3 seconds and then immediately spun at 2000 RMP for 30 sec while a further 1 drop of anhydrous Methyl Acetate was applied. For Layer-by-Layer deposition, a perovskite solution of 50 mg/ml was spin-cast at 2000 RPM for 30 sec. Then the same solid ligand exchanged treatment was applied as in the single step approach for each layer. As a hole-transport layer we used Spiro-OMeTAD (prepared in line with the ref 6) solution which was spin cast under inert atmosphere with a final thickness of 220 nm. Finally, 130 nm of gold electrodes were evaporated with the use of a mask (Pixel area 3.14 mm2).

**Transmission Electron Microscope**: images were obtained using a JEOL JEM-2100 LaB6 transmission electron microscope, operating at 200 kV. The spectrometer is an Oxford Instruments INCA x-sight, with a Si(Li) detector. Samples for TEM characterization were prepared by drop-casting diluted NCs solutions onto 300-mesh carbon-coated copper grids.

**X-ray diffraction**: X-ray diffraction spectra were collected using a PANalytical X'Pert PRO MRD diffractometer equipped with a Cu KαX-ray tube (Cu Kα λ=1.541874Å; Kα$_{1+2}$ doublet λ$_1$=1.540598Å, λ$_2$=1.544426Å). The XRD spectra were collected in air at room temperature using grazing incidence (ω = 0.6º), a parabolic mirror combined with a parallel plate collimator and a parallel-beam geometry. The films used in XRD measurements were prepared by dropcasting the perovskite NCs solution onto a silicon substrate.

**Photoluminescence measurements**: Photoluminescence measurements were performed using a Horiba Jobin Yvon iHR550 Fluorolog system coupled with a Quanta-phi integrating sphere and a FluoroHub time-correlated single photon counting card. Steady-state PL measurements were carried out using a Xenon-lamp coupled with a monocromator as excitation source (λ = 400 nm power density = 1mW/cm$^2$), while for time-resolved measurements a pulsed laser diode was employed (Horiba Nanoled, λ = 405 nm, pulse full-width-half-maximum of 50 ps, fluence of 1nJ/cm$^2$). All photoluminescence efficiency ($\Phi_{PL}$) measurements were carried out in the integrating sphere (λ$_{exc}$ = 520 nm). CsPbBr$_x$I$_{3-x}$ NCs solutions for $\Phi_{PL}$ measurements were prepared in quartz cuvettes and diluted to 0.1 optical density at the excitation wavelength. All films for $\Phi_{PL}$ measurements were prepared via spin-coating at 2000 rpm on soda-lime glass substrates (area of 1 cm$^2$).

**Photovoltaic performance characterization**: solar cells efficiency measurements were performed in the globe box at 25 C°. The device I-V responses were collected using a Keithley 2401 source meter. All the J-V curves were collected with a dwell time of 1ms and a step of 0.02V with the use of a mask. Illumination intensity of AM 1.5 was maintained using a filter and an ABET solar simulator (spectral mismatch within ±20%). The intensity of the solar simulator was calibrated for 1 sun with a standard Si solar cell (provided by Fraunhofer ISE). The measurement and data acquisition was performed with a LabVIEW program.

**EQE measurements:** EQE measurements were performed in air with an in-house build experimental set-up by using chopped (180 Hz, Thorlab) monochromatic illumination. The power of the monochromatic light was measured by a calibrated Newport photo detector (UV-818). The device response of the chopped signal was measured using a Stanford Research system lock-in amplifier (SR-830) which was fed by a Stanford Research system low noise current pre-amplifier (SR-570). The final EQE spectra were obtained with a LabVIEW program.

**RESULTS AND DISCUSSION**

Perovskite NCs of various shapes and sizes can undergo anion exchange reactions extending the degree of control over their optoelectronic response. However, the exchanged iodine based perovskite NCs present poor optical properties, as demonstrated by the lower PLQY compared to their pristine $CsPbBr_3$ counterpart.[21] This arises mainly from structural defects[26] created during the halide replacement reaction, and it limits the applicability of exchanged NCs in optoelectronic devices. In our case, we have employed $CsPbBr_3$ NCs with PLQY up to 83% in solution as a precursor material for the anion exchange with $I^-$. The pristine NCs were synthesized in air and at room temperature, following our previously published procedure.[27] The shape and size of the starting $CsPbBr_3$ NCs is not fully controlled during the synthesis (**Figure 1a**). In line with our synthetic protocol of CsPbBr$_3$ NCs, we performed anion exchange reactions in ambient conditions via dropwise addition of a solution of 230 mg/ml PbI$_2$ in a mixture of octylamine, octanoic acid and hexane (1:1:1 in volume). The exchanged NCs did not show any significant change in morphology (**Figure 1b**). We studied the crystal structure of the mixed halide NCs performing X-ray diffraction measurements (XRD) (**Figure 1c**). The XRD patterns of the mixed halide perovskite NCs indicate a cubic crystal symmetry. We have not observed neither the formation of Cs$_4$PbBr$_x$I$_{6-x}$[28] nor orthorhombic phase (**Figure S1**). The XRD peaks are centered between the CsPbBr$_3$ and CsPbI$_3$ theoretical patterns, thus confirming the mixed halide composition of the NCs (see **Table S1**). We then probed the reaction by collecting the emission spectra of CsPbBr$_x$I$_{3-x}$ NCs with different halide ratios (**Figure 1d**, optical absorption spectra collected in solution are reported in **Figure S2** together with the stoichiometric values obtained from electron energy-dispersive x-ray, EDX, measurements, (see experimental methods). As expected, a red-shift of the emission peak upon increasing the I$^-$/Br$^-$ ratio is observed (from 513 to 676 nm, the full width-half-maximum, FWHM, increase from 25 to 38 nm, see supporting information, **Table S2**). Interestingly, the emission centered at 676 nm corresponds to the CsPbBr$_{0.6}$I$_{2.4}$ NC composition indicating that in the presence of an excess of iodine ions, full bromine replacement is not achieved.

The incomplete exchange may arise from the polydispersity of the initial NCs and their agglomeration, as TEM images manifests (**Figure 1a** and **Figure S3**). The size of the pristine and the exchanged NCs is 14.4 nm and 13 nm respectively while the standard deviation is 7 nm (see **Figure S4** for the size dispersion histograms). Nevertheless, the PL at 676 nm (**Figure 1d**) is comparable to that of pure CsPbI$_3$ NCs (680 nm)[7] indicating that full halide replacement is not necessary to obtain absorption of the VIS spectrum. Indeed, very recently it has been shown that solar cells based on CsPbBrI$_2$ films can reach up to 13% efficiency.[29] We then studied the influence of the NCs concentration in solution on the anion exchange reaction. We focused on the large-scale synthesis of iodide based perovskite which is essential for device fabrication as highly concentrated solutions are required to obtain "pin-hole free" and thick continuous films via spin-coating.

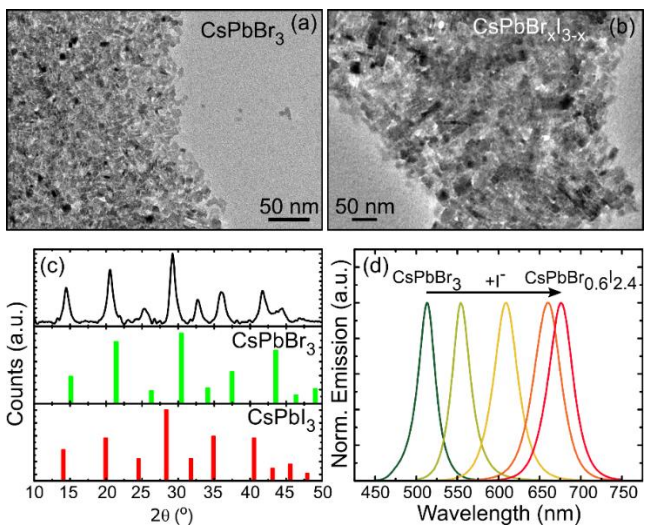

**Figure 1** Bright field transmission electron images of $CsPbBr_3$ (a) and $CsPbBr_xI_{3-x}$ NCs (b), x-ray diffraction pattern of mixed halide NCs with composition $CsPbBrI_2$ (black line), cubic $CsPbBr_3$ (green bars) and cubic $CsPbI_3$ (red bars). (c) Evolution of the photoluminescence spectrum of the $CsPbBrxI_{3-x}$ NCs upon increasing $Br^-$ replacement with $I^-$ in solution (d).

To this end, we carried out the anion exchange reaction on concentrated $CsPbBr_3$ NCs solution (>120 mg/ml) in air. We observed a partial anion exchange in concentrated solutions, resulting in $CsPbBrI_2$ perovskite NCs. Typical absorption and PL spectra of $CsPbBrI_2$ NCs for both film and solution are shown in **Figure 2a**. In the absorption spectra the long Urbach tail is attributed to scattering from small aggregates and the NC size distribution as TEM images suggest. The emission peak at 643 nm in solution, combined with EDX analysis, confirms the formation of $CsPbBrI_2$ NCs. The emission is red shifted in films to 653 nm due to film close-packing which can induce energy transfer from smaller to larger NCs. To assess this hypothesis, we performed spectrally resolved PL decay studies in both film and solution (see **Figure S5**) where the blue-tail of the PL spectrum shows a shorter lifetime than the PL peak, thus confirming the presence of energy transfer. In comparison, perovskite NCs of homogeneous size show spectrally independent PL lifetime.[30] Further studies of the optical response of the exchanged $CsPbBrI_2$ NCs in both film and solution show a record PLQY of 65% in film and 50% in solution.

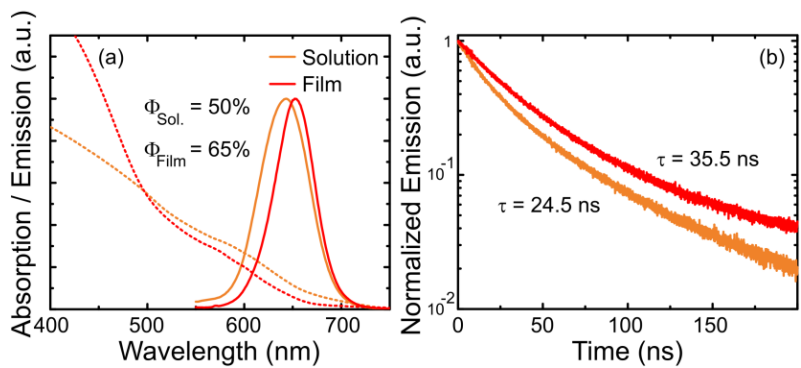

**Figure 2** Optical absorption (dashed line) and photoluminescence (solid line) spectra of the $CsPbBrI_2$ NCs in solution (orange line) and in film (red line), (a) and the respective time resolve PL measurements. Color coding is the same in both panels.

The increase in PLQY upon spin-coating is surprising. Yet, we have previously seen this phenomenon for the pristine $CsPbBr_3$ NCs.[27] A possible explanation is that the PLQY enhancement is due to hole trap passivation from oxygen exposure, as recently been reported.[31,32] The PLQY enhancement is further confirmed by PL time resolved measurement (**Figure 2b**). We have fitted the lifetime traces with mono-exponential decay till the PL intensity drop by 1/e. We measured a PL lifetime ($\tau$) of 35.5 ns and 24.5 ns for both film and solution. We observed that the lifetime increases by 45% in the film which is close to the PLQY enhancement (see **Table S3**).

Considering the good optical properties of the $CsPbBrI_2$ NCs obtained using octanoic acid and octylamine, we tested the use of other shorter surface ligands in the anion-exchange solution (see

**Figure S6**). Reducing the size of the ligands is beneficial for the charge transport of NC films. In fact, transport via charge hopping is known to exponentially depend upon inter-nanocrystal distance.[33,34] We observed that using shorter ligands such as propionic acid and hexanoic acid turned the NCs into the $Cs_4PbBr_xI_{6-x}$ phase within about an hour in ambient conditions (white precipitate in the solution). Only, with the use of octylamine / octanoic acid mixture we do obtain stable nanocrystals.[35] Moreover, we notice that by increasing the length of the primary alkyl group in both carboxylic acid and amine we further enhanced the phase stability of the perovskite NCs in air. Indeed, the use of oleylamine / oleic acid mixture during the anion exchange stabilizes the cubic phase of the NCs for more than a week, as it has been already reported.[7] These results allowed us to fabricate solar cells using two different approaches using NCs with both short and long ligands. First, we employed a single-step (SP) deposition method where the octanoic acid/octylamine capped NCs were dispersed in octane (120 mg/ml) and spin-coated in air, forming a closed packed and dry film of 130 -150 nm. Afterward, we washed away the remaining unbounded ligands from the film´s surface with lead acetate $Pb(OAc)_2$ solution in methyl acetate (MeOAc).[7] In parallel, using the more stable oleic acid/octylamine capped NCs, we built up the film by repeatedly spin casting the NCs (50 mg/ml) from an octane solution, and removed excess ligands using $Pb(OAc)_2$ dissolved in MeOAc,[7] thus obtaining films with a similar thickness to the SP procedure. The cross section of a typical device and the band diagram scheme are shown in **Figure 3a** and **Figure 3b** respectively. For a typical solar cell device we used gold (Au) and fluorine-doped tin oxide (FTO) as hole and electron collecting electrodes, respectively. A thin layer of $TiO_2$ (~ 50 nm) was deposited on top of the FTO glass serving as a hole blocking layer (see **Figure 3b**). Subsequently, a thick layer of Li doped Spiro-OMeTAD (230 nm) was spin-cast on top of the $CsPbBrI_2$ NC film (complete structure: FTO/$TiO_2$/NC film/Spiro-OMETAD/Au).

The J-V scans of both devices are shown in **Figure 3c-d**. Both solar cells present efficiencies up to 5.3 % under 1 sun illumination. The SP device (**Figure 3c**) shows a remarkably high open circuit voltage (Voc) of 1.31 V. To our knowledge, 1.31 V is among the highest $V_{oc}$ reported in perovskite solar cells with band-gap below 2 eV.[36] On the other hand, in the LbL solar cell (**Figure 3d**) the $V_{oc}$ drops down to 1.25 V while the current density increases to 7 mA/cm$^2$, as most of the surface ligands are washed away (see performance summary in Tables S3 and S4). Finally, we collected the J-V curves over time (1000 secs) to study the device stability under 1 sun illumination (**Figure 3f**). We show that the performance of the SP device drops by 10% while the LbL solar cell has a drop of 25%. Nevertheless, the Voc and the the FF in both devices remains relatively stable under continuous illumination (see Figure S7).

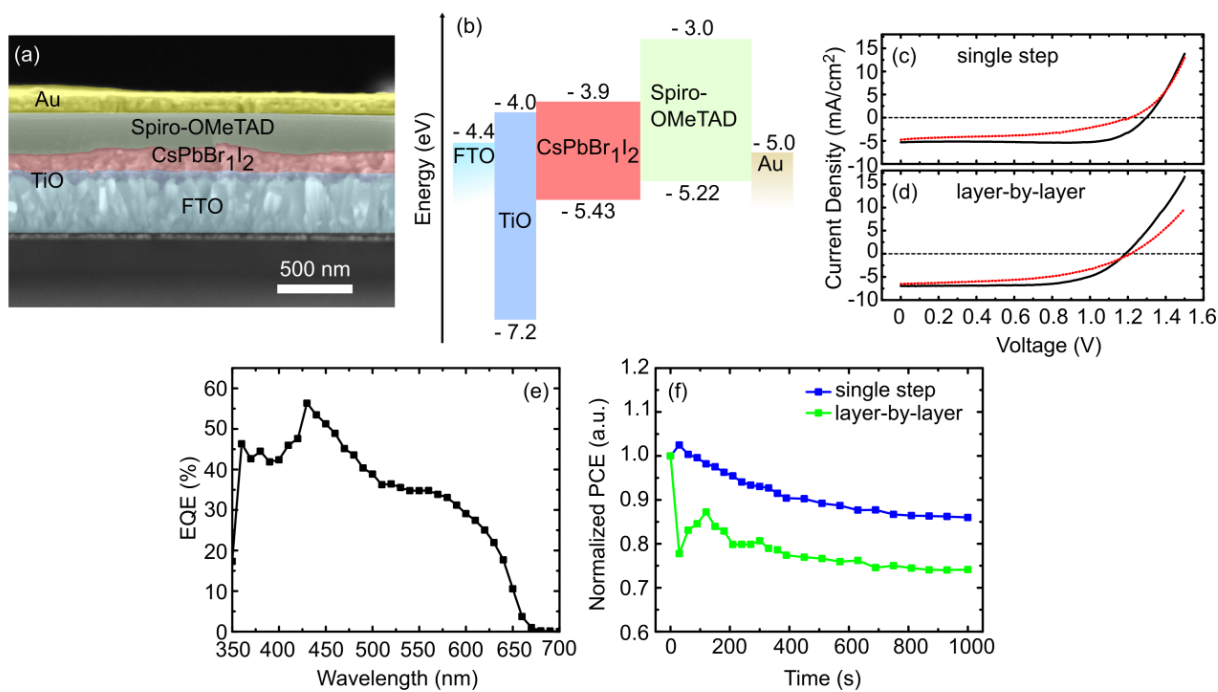

**Figure 3** (a) cross-sectional SEM image of the device (b) schematic of the band diagram of the device. The values have been qualitative extracted from the literature[37] (c)(d) J-V curves of the solar cell devices, reverse scan (black line) and forward scan (red line). (e). Typical EQE spectrum of solar cell device (f) Normalized device power conversion efficiency (PCE) over time under continuous illumination.

Despite the high $V_{oc}$ that we have observed, the external quantum efficiency (EQE) spectrum confirms further that the devices are current limited (**Figure 3e**). The forward scan shows the presence of hysteresis in both devices, as observed and thoroughly studied in bulk perovskite based solar cells[38–40] (see **Table S4**). Nevertheless, the LbL device has suppressed hysteresis in comparison with the SP device, in line with previously reported NC solar cells[7].

**CONCLUSION**

In conclusion, we have synthesized mixed Br/I perovskite nanocrystals in ambient conditions via anion exchange, obtaining a high PLQY of 65% in dry films. We demonstrated a scale-up synthesis of perovskite nanocrystals with mixed halide which does not require high temperature and inert atmosphere to obtain high quality materials. Furthermore, we have shown that by using shorter surface ligands such as octanoic acid, the NCs are phase stable. This allowed us to fabricate a solar cell device via a single step approach with high $V_{oc}$ of 1.31 eV.

**ASSOCIATED CONTENT**

Comparison and analysis of various XRD patterns, TEM images, absorption and PL spectra of NCs with different Br/I ratio, fitting parameters for time resolve PL measurements, time resolve PL maps, size dispersion histograms, NC´s stability with different ligands, summary of device performance and stability .

## AUTHOR INFORMATION

Corresponding Author

*gerasimos.konstantatos@icfo.eu

Author Contributions

‡ S. Christodoulou and F. Di Stasio contributed equally to this work

## ACKNOWLEDGMENT

The authors acknowledge financial support from the European Research Council (ERC) under the European Union's Horizon 2020 research and innovation programme (grant agreement no. 725165), the Spanish Ministry of Economy and Competitiveness (MINECO), and the "Fondo Europeo de Desarrollo Regional" (FEDER) through grant MAT2014-56210-R. Also the authors acknowledge the support from MINECO under the program redes de excellencia TFE. The authors also acknowledge financial support from Fundacio Privada Cellex, the program CERCA and from the Spanish Ministry of Economy and Competitiveness, through the "Severo Ochoa" Programme for Centres of Excellence in R&D (SEV-2015-0522). F. Di Stasio and S. Christodoulou acknowledge support from two Marie Curie Standard European Fellowships ("NANOPTO", H2020-MSCA-IF-2015-703018 and "NAROBAND", H2020-MSCA-IF-2016-750600).

## REFERENCES

(1)   Protesescu, L.; Yakunin, S.; Bodnarchuk, M. I.; Krieg, F.; Caputo, R.; Hendon, C. H.; Yang, R. X.; Walsh, A.; Kovalenko, M. V. Nanocrystals of Cesium Lead Halide Perovskites (CsPbX3, X = Cl, Br, and I): Novel Optoelectronic Materials Showing Bright Emission with Wide Color Gamut. *Nano Lett.* **2015**, *15*, 3692–3696.

(2) Levchuk, I.; Osvet, A.; Tang, X.; Brandl, M.; Perea, J. D.; Hoegl, F.; Matt, G. J.; Hock, R.; Batentschuk, M.; Brabec, C. J. Brightly Luminescent and Color-Tunable Formamidinium Lead Halide Perovskite FAPbX3 (X = Cl, Br, I) Colloidal Nanocrystals. *Nano Lett.* **2017**.

(3) Li, G.; Rivarola, F. W. R.; Davis, N. J. L. K.; Bai, S.; Jellicoe, T. C.; de la Peña, F.; Hou, S.; Ducati, C.; Gao, F.; Friend, R. H.; *et al.* Highly Efficient Perovskite Nanocrystal Light-Emitting Diodes Enabled by a Universal Crosslinking Method. *Adv. Mater.* **2016**, *28*, 3528–3534.

(4) Ling, Y.; Yuan, Z.; Tian, Y.; Wang, X.; Wang, J. C.; Xin, Y.; Hanson, K.; Ma, B.; Gao, H. Bright Light-Emitting Diodes Based on Organometal Halide Perovskite Nanoplatelets. *Adv. Mater.* **2016**, *28*, 305–311.

(5) Zhang, L.; Yang, X.; Jiang, Q.; Wang, P.; Yin, Z.; Zhang, X.; Tan, H.; Yang, Y. (Michael); Wei, M.; Sutherland, B. R.; *et al.* Ultra-Bright and Highly Efficient Inorganic Based Perovskite Light-Emitting Diodes. **2017**, *8*, 15640.

(6) Akkerman, Q. A.; Gandini, M.; Di Stasio, F.; Rastogi, P.; Palazon, F.; Bertoni, G.; Ball, J. M.; Prato, M.; Petrozza, A.; Manna, L. Strongly Emissive Perovskite Nanocrystal Inks for High-Voltage Solar Cells. *Nat. Energy* **2016**, *2*, 16194.

(7) Swarnkar, A.; Marshall, A. R.; Sanehira, E. M.; Chernomordik, B. D.; Moore, D. T.; Christians, J. A.; Chakrabarti, T.; Luther, J. M. Quantum Dot–induced Phase Stabilization of $\alpha$-CsPbI$_3$ Perovskite for High-Efficiency Photovoltaics.

*Science (80-. ).* **2016**, *354*, 92 LP-- 95.

(8)  Palazon, F.; Di Stasio, F.; Akkerman, Q. A.; Krahne, R.; Prato, M.; Manna, L. Polymer-Free Films of Inorganic Halide Perovskite Nanocrystals as UV-to-White Color-Conversion Layers in LEDs. *Chem. Mater.* **2016**.

(9)  Pathak, S.; Sakai, N.; Wisnivesky Rocca Rivarola, F.; Stranks, S. D.; Liu, J.; Eperon, G. E.; Ducati, C.; Wojciechowski, K.; Griffiths, J. T.; Haghighirad, A. A.; *et al.* Perovskite Crystals for Tunable White Light Emission. *Chem. Mater.* **2015**, *27*, 8066–8075.

(10) Zhao, H.; Zhou, Y.; Benetti, D.; Ma, D.; Rosei, F. Perovskite Quantum Dots Integrated in Large-Area Luminescent Solar Concentrators. *Nano Energy* **2017**, *37*, 214–223.

(11) Polavarapu, L.; Nickel, B.; Feldmann, J.; Urban, A. S. Advances in Quantum-Confined Perovskite Nanocrystals for Optoelectronics. *Adv. Energy Mater.* **2017**, 1700267.

(12) Protesescu, L.; Yakunin, S.; Kumar, S.; Bär, J.; Bertolotti, F.; Masciocchi, N.; Guagliardi, A.; Grotevent, M.; Shorubalko, I.; Bodnarchuk, M. I.; *et al.* Dismantling the "Red Wall" of Colloidal Perovskites: Highly Luminescent Formamidinium and Formamidinium–Cesium Lead Iodide Nanocrystals. *ACS Nano* **2017**, *11*, 3119–3134.

(13) Imran, M.; Di Stasio, F.; Dang, Z.; Canale, C.; Khan, A. H.; Shamsi, J.; Brescia, R.; Prato, M.; Manna, L. Colloidal Synthesis of Strongly Fluorescent CsPbBr3 Nanowires with Width Tunable Down to the Quantum Confinement Regime. *Chem. Mater.* **2016**, *28*, 6450–6454.

(14) Zhang, D.; Yu, Y.; Bekenstein, Y.; Wong, A. B.; Alivisatos, A. P.; Yang, P. Ultrathin Colloidal Cesium Lead Halide Perovskite Nanowires. *J. Am. Chem. Soc.* **2016**, *138*, 13155–13158.

(15) Akkerman, Q. A.; Motti, S. G.; Srimath Kandada, A. R.; Mosconi, E.; D'Innocenzo, V.; Bertoni, G.; Marras, S.; Kamino, B. A.; Miranda, L.; De Angelis, F.; *et al.* Solution Synthesis Approach to Colloidal Cesium Lead Halide Perovskite Nanoplatelets with Monolayer-Level Thickness Control. *J. Am. Chem. Soc.* **2016**, *138*, 1010–1016.

(16) Bekenstein, Y.; Koscher, B. A.; Eaton, S. W.; Yang, P.; Alivisatos, A. P. Highly Luminescent Colloidal Nanoplates of Perovskite Cesium Lead Halide and Their Oriented Assemblies. *J. Am. Chem. Soc.* **2015**, *137*, 16008–16011.

(17) Yang, W. S.; Park, B.-W.; Jung, E. H.; Jeon, N. J.; Kim, Y. C.; Lee, D. U.; Shin, S. S.; Seo, J.; Kim, E. K.; Noh, J. H.; *et al.* Iodide Management in Formamidinium-Lead-Halide–based Perovskite Layers for Efficient Solar Cells. *Science (80-. ).* **2017**, *356*.

(18) McMeekin, D. P.; Sadoughi, G.; Rehman, W.; Eperon, G. E.; Saliba, M.; Hörantner, M. T.; Haghighirad, A.; Sakai, N.; Korte, L.; Rech, B.; *et al.* A Mixed-Cation Lead Mixed-Halide Perovskite Absorber for Tandem Solar Cells. *Science (80-. ).* **2016**, *351*.

(19) Lee, J.-W.; Hsieh, Y.-T.; De Marco, N.; Bae, S.-H.; Han, Q.; Yang, Y. Halide Perovskites for Tandem Solar Cells. *J. Phys. Chem. Lett.* **2017**, *8*, 1999–2011.

(20) Bush, K. A.; Palmstrom, A. F.; Yu, Z. J.; Boccard, M.; Cheacharoen, R.; Mailoa, J. P.; McMeekin, D. P.; Hoye, R. L. Z.; Bailie, C. D.; Leijtens, T.; *et al.* 23.6%-Efficient Monolithic Perovskite/silicon Tandem Solar Cells with Improved Stability. **2017**, *2*, 17009.

(21) Akkerman, Q. A.; D'Innocenzo, V.; Accornero, S.; Scarpellini, A.; Petrozza, A.; Prato, M.; Manna, L. Tuning the Optical Properties of Cesium Lead Halide Perovskite Nanocrystals by Anion Exchange Reactions. *J. Am. Chem. Soc.* **2015**, *137*, 10276–10281.

(22) Zhang, T.; Li, G.; Chang, Y.; Wang, X.; Zhang, B.; Mou, H.; Jiang, Y.; Zhou, H.; Chen, L.; Zhong, H.; *et al.* Full-Spectra Hyperfluorescence Cesium Lead Halide Perovskite Nanocrystals Obtained by Efficient Halogen Anion Exchange Using Zinc Halogenide Salts. *CrystEngComm* **2017**, *19*, 1165–1171.

(23) Nedelcu, G.; Protesescu, L.; Yakunin, S.; Bodnarchuk, M. I.; Grotevent, M. J.; Kovalenko, M. V. Fast Anion-Exchange in Highly Luminescent Nanocrystals of Cesium Lead Halide Perovskites (CsPbX3, X = Cl, Br, I). *Nano Lett.* **2015**, *15*, 5635–5640.

(24) Tong, Y.; Bladt, E.; Aygüler, M. F.; Manzi, A.; Milowska, K. Z.; Hintermayr, V. A.; Docampo, P.; Bals, S.; Urban, A. S.; Polavarapu, L.; *et al.* Highly Luminescent Cesium Lead Halide Perovskite Nanocrystals with Tunable Composition and Thickness by Ultrasonication. *Angew. Chemie Int. Ed.* **2016**, *55*, 13887–13892.

(25) Li, X.; Wu, Y.; Zhang, S.; Cai, B.; Gu, Y.; Song, J.; Zeng, H. CsPbX$_3$ Quantum Dots for Lighting and Displays: Room-Temperature Synthesis, Photoluminescence Superiorities, Underlying Origins and White Light-Emitting Diodes. *Adv. Funct. Mater.* **2016**, *26*, 2435–2445.

(26) Ball, James M.; Petrozza, A. Defects in Perovskite-Halides and Their Effects in Solar Cells. *Nano Energy* **2016**, *1*, 16149.

(27) Stasio, F. Di; Christodoulou, S.; Huo, N.; Konstantatos, G.; Di Stasio, F.; Christodoulou, S.; Huo, N.; Konstantatos, G. Near-Unity Photoluminescence Quantum Yield in CsPbBr3 Nanocrystal Solid-State Films via Post-Synthesis Treatment with Lead Bromide. *Chem. Mater.* **2017**, *29*, acs.chemmater.7b02834.

(28) Akkerman, Q. A.; Park, S.; Radicchi, E.; Nunzi, F.; Mosconi, E.; De Angelis, F.; Brescia, R.; Rastogi, P.; Prato, M.; Manna, L. Nearly Monodisperse Insulator Cs4PbX6 (X = Cl, Br, I) Nanocrystals, Their Mixed Halide Compositions, and Their Transformation into CsPbX3 Nanocrystals. *Nano Lett.* **2017**, *17*, 1924–1930.

(29) Liu, C.; Li, W.; Zhang, C.; Ma, Y.; Fan, J.; Mai, Y. All-Inorganic CsPbI$_2$Br Perovskite Solar Cells with High Efficiency Exceeding 13%. *J. Am. Chem. Soc.* **2018**, jacs.7b13229.

(30) Swarnkar, A.; Chulliyil, R.; Ravi, V. K.; Irfanullah, M.; Chowdhury, A.; Nag, A. Colloidal CsPbBr$_3$ Perovskite Nanocrystals: Luminescence beyond Traditional Quantum Dots. *Angew. Chemie* **2015**, *127*, 15644–15648.

(31) Lorenzon, M.; Sortino, L.; Akkerman, Q.; Accornero, S.; Pedrini, J.; Prato, M.; Pinchetti, V.; Meinardi, F.; Manna, L.; Brovelli, S. Role of Nonradiative Defects and Environmental Oxygen on Exciton Recombination Processes in CsPbBr$_3$ Perovskite Nanocrystals. *Nano Lett.* **2017**, *17*, 3844–3853.

(32) Meggiolaro, D.; Mosconi, E.; De Angelis, F. Mechanism of Reversible Trap Passivation by Molecular Oxygen in Lead-Halide Perovskites. *ACS Energy Lett.* **2017**, *2*, 2794–2798.

(33) Kodaimati, M. S.; Wang, C.; Chapman, C.; Schatz, G. C.; Weiss, E. A. Distance-Dependence of Interparticle Energy Transfer in the Near-Infrared within Electrostatic Assemblies of PbS Quantum Dots. *ACS Nano* **2017**, *11*, 5041–5050.

(34) Liu, Y.; Gibbs, M.; Puthussery, J.; Gaik, S.; Ihly, R.; Hillhouse, H. W.; Law, M. Dependence of Carrier Mobility on Nanocrystal Size and Ligand Length in PbSe Nanocrystal Solids. *Nano Lett.* **2010**, *10*, 1960–1969.

(35) Yang, J.; Siempelkamp, B. D.; Liu, D.; Kelly, T. L. Investigation of $CH_3NH_3PbI_3$ Degradation Rates and Mechanisms in Controlled Humidity Environments Using *in Situ* Techniques. *ACS Nano* **2015**, *9*, 1955–1963.

(36) Niezgoda, J. S.; Foley, B. J.; Chen, A. Z.; Choi, J. J. Improved Charge Collection in Highly Efficient $CsPbBrI_2$ Solar Cells with Light-Induced Dealloying. *ACS Energy Lett.* **2017**, *2*, 1043–1049.

(37) Sidhik, S.; Esparza, D.; Martínez-Benítez, A.; Lopez-Luke, T.; Carriles, R.; Mora-Sero, I.; de la Rosa, E. Enhanced Photovoltaic Performance of Mesoscopic Perovskite Solar Cells by Controlling the Interaction between $CH_3NH_3PbI_3$ Films and $CsPbX_3$ Perovskite Nanoparticles. *J. Phys. Chem. C* **2017**, *121*, 4239–4245.

(38) Wu, Y.; Shen, H.; Walter, D.; Jacobs, D.; Duong, T.; Peng, J.; Jiang, L.; Cheng, Y.-B.; Weber, K. On the Origin of Hysteresis in Perovskite Solar Cells. *Adv. Funct. Mater.* **2016**, *26*, 6807–6813.

(39) Chen, B.; Yang, M.; Priya, S.; Zhu, K. Origin of $J-V$ Hysteresis in Perovskite Solar Cells. *J. Phys. Chem. Lett.* **2016**, *7*, 905–917.

(40) Nemnes, G. A.; Besleaga, C.; Stancu, V.; Dogaru, D. E.; Leonat, L. N.; Pintilie, L.; Torfason, K.; Ilkov, M.; Manolescu, A.; Pintilie, I. Normal and Inverted Hysteresis in Perovskite Solar Cells. *J. Phys. Chem. C* **2017**, *121*, 11207–11214.